\documentclass[aps,prb,twocolumn,superscriptaddress,longbibliography]{revtex4-2}
\usepackage{times,bm,soul,color}
\usepackage{amssymb,amsmath,amsfonts,mathrsfs,physics,booktabs}
\usepackage{graphicx,dcolumn,multirow}
\usepackage[svgnames]{xcolor}
\usepackage[colorlinks,bookmarks=false,citecolor=blue,linkcolor=red,urlcolor=blue]{hyperref}
\usepackage{subfigure}
\usepackage{anyfontsize}
\usepackage{blkarray}
\usepackage[percent]{overpic}

\renewcommand{\Re}{\operatorname{Re}}
\renewcommand{\Im}{\operatorname{Im}}

\begin{document}

\title{Non-Bloch Dirac Points and Phase Diagram in the Stacked Non-Hermitian SSH Model}

\author{Megan Schoenzeit}
\affiliation{School of Physics and Astronomy,
University of Minnesota, Minneapolis, MN 55455, United States}
\author{Chang Shu}
\author{Kai Zhang}
\author{Kai Sun}
\affiliation{Department of Physics, University of Michigan, Ann Arbor, MI 48109, United States}

\begin{abstract}

Topological semimetals exhibit protected band crossings in momentum space, accompanied by corresponding surface states. 
Non-Hermitian Hamiltonians introduce geometry-sensitive features that dissolve this bulk-boundary correspondence principle. 
In this paper, we exemplify this phenomenon by investigating a non-Hermitian 2D stacked SSH chain model with non-reciprocal hopping and on-site gain/loss. 
We derive an analytical phase diagram in terms of the complex energy gaps in the open-boundary spectrum. 
The phase diagram reveals the existence of non-Bloch Dirac points, which feature a real spectrum and only appear under open boundary conditions but disappear in Bloch bands under periodic boundary conditions.  
Due to the reality of the spectrum in the vicinity of non-Bloch Dirac points, we can locally map it to Hermitian semimetals within the Altland-Zirnabuer symmetry classes. 
Based on this mapping, we demonstrate that non-Bloch Dirac points are characterized by an integer topological charge. 
Unlike the band crossings in Hermitian semimetals, the locations of the non-Bloch Dirac points under different boundary geometries do not match each other, indicating a geometry-dependent bulk-boundary correspondence in non-Hermitian semimetals. 
Our findings provide new pathways into establishing unconventional bulk-boundary correspondence for non-Bloch Dirac metals in non-Hermitian systems. 
\end{abstract}

\maketitle
\section{Introduction}
Bulk-boundary correspondence (BBC) is a fundamental principle of topological band theory, where topological edge states robust against perturbations are protected by the topologically non-trivial band structure of the bulk~\cite{Kane2005,BHZ2006,Schnyder2008,Qi2011Review,Fu2007,Chiu2016}. 
For instance, in the study of topological semimetals, e.g. Dirac semimetals in 2D and Weyl semimetals in 3D, the topological invariants can be calculated on a surface surrounding these band crossings. 
Correspondingly, the Fermi-arc states terminating at the band crossings, projected onto the edge, can be observed, which serves as a hallmark of bulk-boundary correspondence in topological semimetals~\cite{Ryu2002,Armitage2018,gao2019topological,lv2021experimental}. 
For Hermitian band systems, the energy spectra and eigenstates in the bulk are insensitive to boundary conditions. Attributing to this insensitivity, the knowledge of Bloch band crossings, e.g., Dirac and Weyl nodes, is sufficient to capture all information of topological edge states under various open boundary geometries in Hermitian semimetals. 

Nevertheless, for a class of non-Hermitian systems, the system's spectrum and eigenstates are notoriously sensitive to boundary conditions, known as non-Hermitian skin effect~\cite{Lee2016,Yao2018,Kunst2018,Murakami2019PRL,ChingHua2019,Slager2020PRL,Zhang2020,Okuma2020,kawabata2019,zhang2022universal,XJLiu2023PRB,LeeCHReview}. 
Thus, the conventional BBC fails in these non-Hermitian systems, as the bulk topological invariants derived from Bloch bands do not align with the emergence of edge states. Instead, accurately characterizing the edge states requires the open-boundary continuum spectrum, which is determined through a complex analytical extension of the Brillouin zone, known as the non-Bloch formulation~\cite{Yao2018,WangZhong2018,Murakami2019PRL,ZSaGBZPRL,KawabataPRB2020,Murakami2022,HYWang2022}. 
Building on the well-established one-dimensional non-Bloch band framework, substantial research has been conducted on the 1D non-Bloch BBC, encompassing both gapped and gapless topological phases~\cite{Yokomizo2020PRR,xiao2020non,Yang2020,DengTS2019,okuma2023}. 
However, the study of non-Bloch semimetals in two and higher dimensions remains limited. 
Although non-Hermitian Dirac and Weyl points have been investigated~\cite{Xue2020}, these band crossings are defined in systems with PBCs and deviates far from open boundary systems, especially when the non-Hermitian skin effect is present.  
Furthermore, in two and higher dimensions, the system's spectrum strongly depends on its geometry, a phenomenon known as the geometry-dependent skin effect~\cite{Kai2022NC,Wang2022NC,DingKun2023PRL,QYZhou2023NC,WanTuoSciB,QinYi2023PRA,GBJo2023arXiv}. 
This interplay between non-Bloch band crossings and geometric effects would reshape the topological characterization of non-Bloch semimetals and remains an open area of study. 

In this work, we study a 2D non-Hermitian model composed of coupled Su-Schrieffer-Heeger (SSH) chains~\cite{ssh1979,Lieu2018,Yuce2019,Wu2021}.
We managed to compute the edge spectrum analytically and subsequently derive the analytical phase diagram.
The non-Bloch Dirac points (DPs) show up at one of the phase transition points between two non-Bloch $\mathcal{PT}$ exact phases with a real spectrum. 
We further study the topological properties of the non-Bloch Dirac points, which are OBC continuum band crossings characterized by linear dispersion and a locally real spectrum. 
Utilizing the reality of the spectrum near the non-Bloch DP, we can locally map it to a Hermitian DP by an imaginary gauge transformation. Then, according to the classification scheme in Hermitian topological semimetals, we find that the non-Bloch DP is featured by a $\mathbb{Z}$ topological charge, with the chiral and/or mirror symmetry~\cite{Chiu2016,Chiu2014classification}. 
Notably, when projecting the non-Bloch band structure to edges with different tilting angles, the locations of the DPs are not consistent with each other, totally in disagreement with the doctrines of BBC in Hermitian semimetals. 
Our work sheds light on the topological features of the non-Bloch DPs under open boundary in 2D and may inspire further investigation along this line. 

The rest of the paper is organized as follows:
In Sec.~\ref{sec:phase_diagram}, we introduce the 2D non-Hermitian stacked SSH chain model and analytically derive its phase diagram in terms of the energy line gaps. We also demonstrate the key features of the energy spectrum of each phase.
In Sec.~\ref{sec:DPs}, we investigate the topological properties of the non-Bloch DPs. In Sec. ~\ref{sec:classification}, we map the Hamiltonian to a Hermitian problem and show the non-Bloch DPs are protected by a $\mathbb{Z}$ index.
In Sec.~\ref{sec:generalization}, we generalize our model by introducing next-nearest neighbor hoppings to the model and studying the symmetry-protected stability of the DPs.
In Sec.~\ref{sec:OBCPBCdistinction}, we show the breakdown of the conventional bulk-boundary correspondence.
Specifically, the DPs are absent in the PBC spectrum.  
In Sec.~\ref{sec:conclusion}, we summarize our results and provide discussions and outlooks.

\section{Phase Diagram Analytic Solution}\label{sec:phase_diagram}
\begin{figure}[ht!]
\centering
\includegraphics[width=1\linewidth]{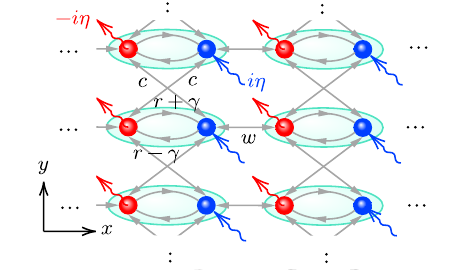}
\caption{Graphical representation of 2D stacked SSH model with sublattice sites A (red) and B (blue). The green ovals represent unit cells, the gray arrows represent inter-/intra-cell couplings, and the arrows with tildes represent mode gain/loss. 
All the parameters $(r,\gamma,c,w,\eta)$ have been marked in the diagram.
}
\label{fig1}
\end{figure}

We begin by introducing the Hamiltonian of the non-Hermitian 2D stacked SSH chain model.
The diagrammatic illustration of this model is shown in Fig.~\ref{fig1}.
As demonstrated below, this model exhibits dramatic changes to the spectrum and eigenmodes when open boundaries are introduced, compared to those under periodic boundary conditions. Remarkably, the model remains analytically solvable even with open boundaries, allowing us to derive the exact phase diagram in this setting. This analytical framework enables a complete characterization of the phase diagram, the precise identification of non-Hermitian nodal points, and a detailed analysis of their topological properties. 

The Hamiltonian in momentum space is given by $\hat{H}=\sum_{\mathbf{k}}\hat{\psi}^\dagger_{\mathbf{k}} \mathcal{H}(\mathbf{k}) \hat{\psi}_{\mathbf{k}}$ where $\mathbf{k}=(k_x,k_y)$ and $\hat{\psi}^{\dagger}_{\mathbf{k}}=(\hat{c}_{A,\mathbf{k}}^\dagger,\hat{c}_{B,\mathbf{k}}^\dagger)$.
Here $\hat{c}_{A,\mathbf{k}}^\dagger$ and $\hat{c}_{B,\mathbf{k}}^\dagger$ denote the fermion creation operators of sublattice A and B, respectively, as indicated by the red and blue dots in Fig.~\ref{fig1}.
The Hamiltonian matrix is given by
\begin{align}\label{eq:Hk}
    \mathcal{H}(k_x,k_y)&=(i\gamma-w\sin k_x)\sigma_x\notag\\
    &+(r+2c\cos k_y+w\cos k_x)\sigma_y+i\eta\sigma_z
\end{align}
where $\sigma_x,\sigma_y,$ and $\sigma_z$ are Pauli matrices.

Here $r+\gamma$ and $r-\gamma$ are asymmetric intra-cell hopping while $w$ is the inter-cell hopping.
$c$ is the off-diagonal coupling along $y$, and $\eta$ represents the strength of onsite gain and loss. All parameters are assumed to be real. 
This model could be experimentally implemented in mechanical/acoustic metamaterials, or exciton/polariton systems consisting of a heterostructure stacked by photonic crystals and 2D materials. 

Since the hopping along the $y$ direction is Hermitian, taking PBC and OBC in $y$ will yield the same results.
To analytically obtain the fully open boundary spectrum, we take PBC in $y$ and OBC in $x$, thus forming the cylinder geometry.
To incorporate the OBC in $x$, we use complex-valued non-Bloch wave vector $\beta_x:=e^{ik_x}$, which takes value in the Generalized Brillouin Zone (GBZ)~\cite{Yao2018}.
Due to translational symmetry in $y$, we have Bloch momentum $k_{y}\in[-\pi,\pi)$. 
Let $t = r + 2c \cos{(k_y)}$. Plugging this into Eq.~\eqref{eq:Hk} and converting to polynomials in $\beta_x$ yields 
\begin{equation}\label{eq2}
\widetilde{\mathcal{H}}(\beta_x, k_y)=\mqty(i\eta & t + \gamma + w\beta _{x} \\ 
t - \gamma + w/\beta _{x} & -i\eta).
\end{equation}
We now solve the characteristic equation $f(E, \beta_x, k_y)=\operatorname{det}[E\mathbb{I}_2-\widetilde{\mathcal{H}}(\beta_x, k_y)]=0$ and obtain 
\begin{equation}\label{eq3}
E^2 = (t-\gamma) w\beta_{x}  + \frac{w(t+\gamma)}{\beta_{x}}+t^2 + w^2-\gamma^2-\eta^2.
\end{equation}
The solutions for $\beta_x$ as a function of $E$ are
\begin{equation}\label{eq4}
\beta_{x,\pm} = \frac{\sqrt{\gamma^2-t^2}}{t-\gamma}\left(v\pm\sqrt{1+v^2}\right)
\end{equation}
where 
\begin{equation}\label{eq5}
v=\frac{E^2-t^2-w^2+\gamma^2+\eta^2}{2w\sqrt{\gamma^2-t^2}}.
\end{equation}
To obey the standing wave condition~\cite{Yao2018,Murakami2019PRL} along $x$, the two $\beta_x$ solutions must have the same amplitude:
\begin{equation}\label{eq6}
\abs{\beta_{x,+}} = \abs{\beta_{x,-}}.
\end{equation}
Hence, from Eq.~\eqref{eq4}, we conclude that $\abs{v+\sqrt{1+v^2}}=\abs{v-\sqrt{1+v^2}}$. 
Due to the identity $(v+\sqrt{1+v^2})(v-\sqrt{1+v^2})=-1$, we must have $\abs{ v+\sqrt{1+v^2}}=1$. 
Being free to take any phase factor, we set $v+\sqrt{1+v^2} = e^{i\phi}$ with $\phi\in[-\pi,\pi)$.
Solving for $v$ we get $v = i\sin{\phi}$.
Now, plugging this back into Eq.~\eqref{eq4} and solving for $E$ yields
\begin{equation}\label{eq7}
E = \pm \sqrt{t^2 + w^2 -\gamma^2 -\eta^2 + 2w\sin{\phi}\sqrt{t^2 - \gamma^2}} 
\end{equation}
Notice that $t=t(k_y)$. By sweeping through $k_y$ and $\phi$, the energy spectrum of the model can be obtained.

\begin{figure*}[t]
\begin{center}
\includegraphics[width=\linewidth]{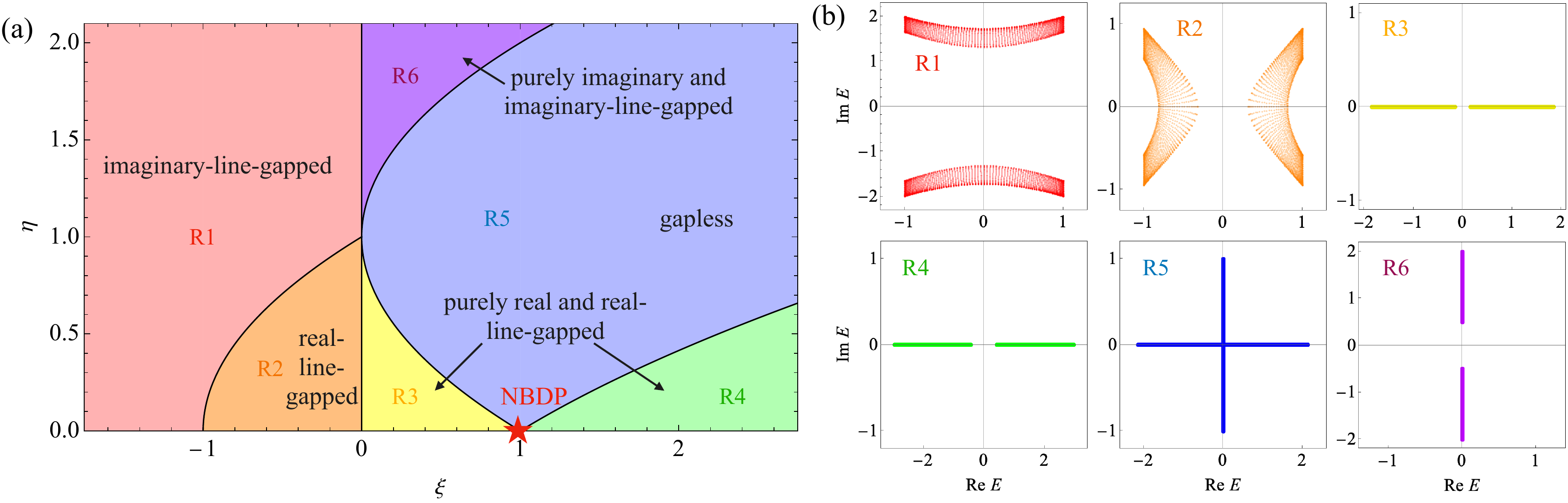}
\caption{Phase diagram (left image) and open boundary energy spectra for each phase (right panels). Only the $\eta>0$ part is displayed as the phase diagram is symmetric about the $\eta = 0$. Crossing the star at $(\xi,\eta) = (1,0)$ creates non-Bloch Dirac points, which will be discussed in detail in Sec.~\ref{sec:DPs}. The energy spectra for R1-R3 are in the top row (left to right) and for R4-R6 are in the bottom row (left to right). Regions R1 and R2 are defined for $\xi <0 $ and have both real and imaginary energy parts. Regions R3-R6 are defined for $\xi>0$ and the entire energy spectrum collapses into the real and imaginary axes.} 
\label{fig2}
\end{center}
\end{figure*} 

The same result can be obtained using the auxiliary GBZ method~\cite{Yang2020}.
The cylinder GBZ can be obtained as a 2D surface embedded in the 3D space of $(\Re \beta_x,\Im\beta_x,k_y)$.
For each $k_y\in[-\pi,\pi)$, the intersection between GBZ and the constant $k_y$ plane is a circle with radius $R(k_y):=|\beta_x(k_y)|$ given by
\begin{equation}\label{eq8}
    R(k_y)=\sqrt{\abs{\frac{t(k_y)+\gamma}{t(k_y)-\gamma}}}
\end{equation}
Plugging $\beta_x(k_y)=R(k_y)e^{i\phi}$ into Eq.~\eqref{eq2}, we can also get the energy spectrum Eq.~\eqref{eq7}.
The spectrum can be entirely real, entirely imaginary, or a mix of real and imaginary components, depending on the sign of the expression within the square root in Eq.~\eqref{eq7}. 
In the remainder of this section, we will elucidate the phase diagram for the energy spectrum of the model, based on the sign of $\xi := t^2 - \gamma^2$. 
For simplicity, we set $w = 1$ without loss of generality since we can scale the total energy. 
Additionally, we assume $\eta > 0$ as the negative case yields essentially the same results.

\vspace{1em}
\paragraph{Case I: $\xi\geq0$}
The expression inside the square root in Eq.~\eqref{eq7} is real when $\xi\geq0$.
This indicates that the entire spectrum is confined to the real and imaginary axes and does not cover a finite area on the complex plane.
Notice that $E^2$ reaches its maximum and minimum when $\sin\phi=1$ and $\sin\phi=-1$, respectively. 
Therefore,
\begin{equation}\label{eq9}
        (1-\sqrt{\xi})^2-\eta^2 < E^2 < (1+\sqrt{\xi})^2-\eta^2
\end{equation}
The entire spectrum is real if $(1-\sqrt{\xi})^2-\eta^2\geq 0$.
Similarly, the entire spectrum is imaginary if $(1+\sqrt{\xi})^2-\eta^2\leq 0$.
Otherwise, if $(1-\sqrt{\xi})^2-\eta^2< 0<(1+\sqrt{\xi})^2-\eta^2$, the spectrum partially lies on the real axis and partially lies on the imaginary axis.
So with $\xi>0$ we get two phase boundaries $\eta=\sqrt{\xi}-1$ with $\xi>1$ and $\xi=(\eta-1)^2$ with $\eta>0$.

\vspace{1em}
\paragraph{Case II: $\xi<0$} When $\xi<0$, Eq.~\eqref{eq7} can be cast into
\begin{equation}\label{eq10}
    E = \pm \sqrt{1+\xi -\eta^2 + 2i\sin{\phi}\sqrt{|\xi|}} 
\end{equation}
When $1+\xi-\eta^2<0$, the real part inside the square root is negative, and thus $E$ can never be purely real.
Therefore, the spectrum in this case never touches the real axis, i.e. has a real line gap.
Similarly, when $1+\xi-\eta^2>0$, the real part inside the square root is positive, so $E$ cannot be purely imaginary and thus has an imaginary line gap. 
Hence, the phase boundary is given by $1+\xi-\eta^2=0$ or equivalently $\eta=\sqrt{1+\xi}$ with $-1<\xi<0$.

From the discussion above, we obtain the phase diagram of our model in the $(\xi,\eta)$ plane displayed in Fig.~\ref{fig2}.
There are 4 phase boundaries given by
\begin{enumerate}
    \item $\xi=0$;
    \item $\eta=\sqrt{\xi+1}$ where $-1<\xi<0$;
    \item $\eta=\sqrt{\xi}-1$ where $\xi>1$;
    \item $\xi=(\eta-1)^2$ where $\eta>0$.
\end{enumerate}
These phase boundaries partition the $(\xi, \eta)$ plane into six distinct regions R1-R6, each representing a phase with unique spectral characteristics.

Next, we outline the key characteristics of each phase in Fig.~\ref{fig2} and the transitions between them. 
According to the previous analysis, phase R1 is imaginary-line-gapped and R2 is real-line-gapped. 
When $\xi>0$, however, the spectrum collapses entirely onto the real/imaginary axes. 
In both phases R3 and R4, the spectrum lies on the real axis, with both phases being gapped by the imaginary axis. 
Phase R5 is gapless with both real and imaginary parts of the spectrum intersecting at the origin.
Phase R6 is purely imaginary and is gapped by the real axis.
The energy spectra for these six phases are shown in Fig.~\ref{fig2} (b).

Since the band structure is generated by taking $k_y\in[-\pi,\pi)$ while the horizontal coordinate of the phase diagram $\xi$ depends on $k_y$, we may encounter phase boundaries in the band structure. 
These numerous crossings exhibit a rich variety of behaviors, comprising the merging and separation of real and imaginary line-gapped bands, as well as the transitions between real line gaps and imaginary line gaps.
The band structure contains two notable types of special $k_y$ points in the band structure: the EPs and the non-Bloch Dirac points.
A notable example is the R2-R3-R4 crossing in the phase diagram with $\eta{=}0$, which features both exceptional points (EPs) and non-Bloch DPs. 
In the R2-R3 transition, the spectrum collapses to the real axis, indicating that the phase boundary between R2 and R3 is an EP.
Subsequently, in the R3-R4 transition, the real line gap first closes and then reopens. The point at which the gap closes is precisely the non-Bloch DP, denoted by the red asterisk in the phase diagram Fig.~\ref{fig2} (a).

Here, we use the term \emph{non-Bloch} to signify the gapless points in the open boundary spectrum.
This is in sharp contrast with the previous study of non-Hermitian Dirac cones and other band crossings~\cite{Xue2020}, where they focus on the Bloch spectrum under PBC.

\section{Non-Bloch Dirac Points}\label{sec:DPs}

The conventional Dirac points in the Bloch spectrum indicate a linear band dispersion near the band crossing points.
To justify our terminology, we need to redefine a 2D parameter space when studying the OBC spectrum, in our case, spanned by $k_y$ and $\phi$ as given by Eq.~\eqref{eq7}.
Then, we need to verify that the local dispersion near the non-Bloch DP is linear.

\begin{figure}[ht]
\centering
\includegraphics[width=\linewidth]{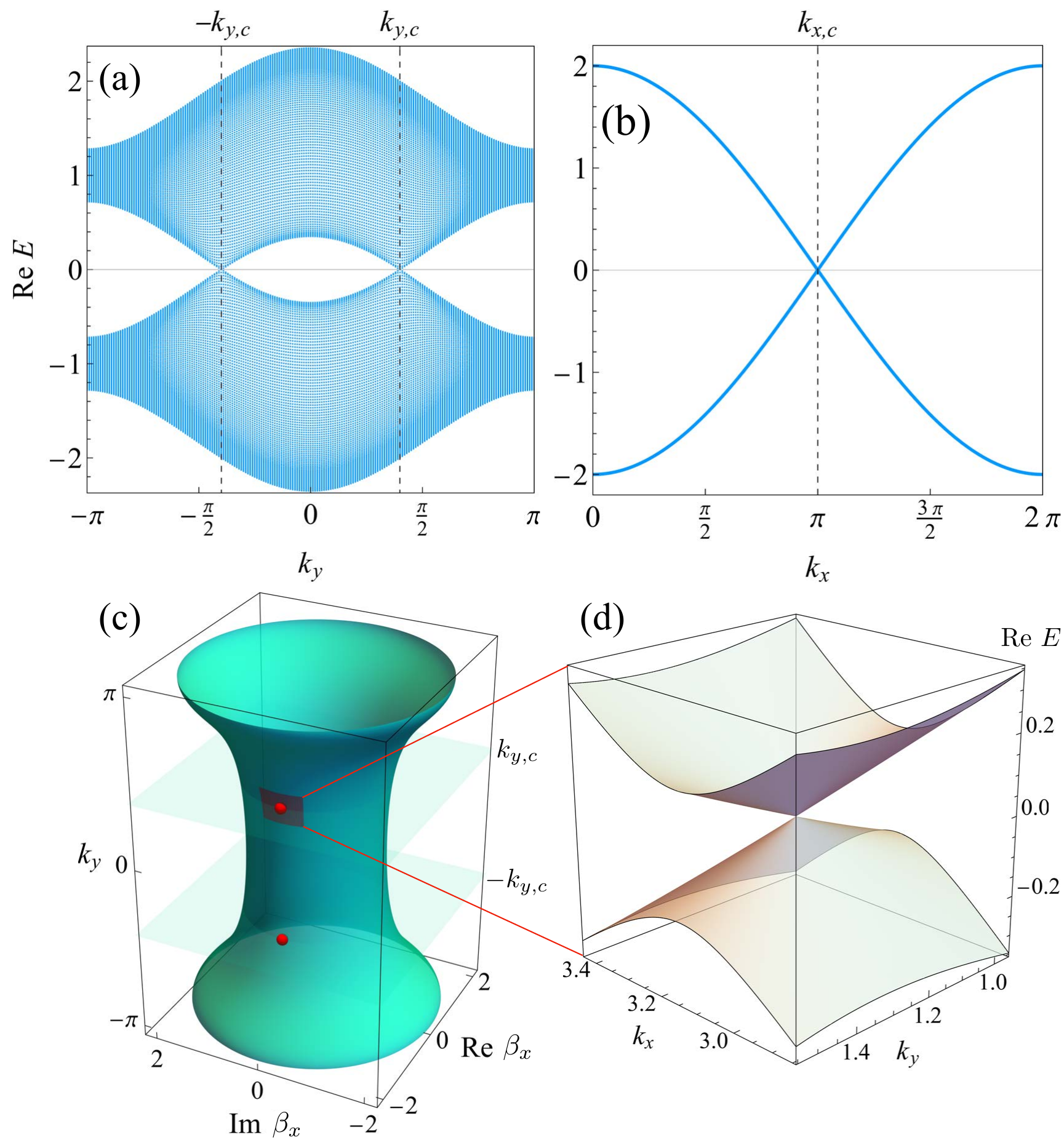}
\caption{The non-Bloch Dirac points in the band structure. Here we take $(r,c,\gamma,w,\eta)=(\frac{7}{8},\frac{1}{4},\frac{1}{4},1,0)$ such that we go from phase R3 to R4 by crossing the non-Bloch DP.
(a) shows the non-Bloch band structure of the system as a function of $k_y$.
There are two gap-closing points at $\pm k_{y,c}$.
(b) Fixing $k_y=k_{y,c}$, we can plot the band structure as a function of $\theta_x$. The non-Bloch DP is at $\theta_x=\pi$.
(c) The cylinder GBZ surface, with two non-Bloch DPs marked in red.
(d) is the local magnification of the non-Bloch band structure near the non-Bloch DP.
    }
\label{fig3}
\end{figure}

We may view the system as a quasi-1D ribbon along $y$ and plot the band structure as a function of $k_y$.
In crossing regions R3-R4 with $\eta{=}0$, the OBC spectrum is entirely real.
The phase boundary between R3 and R4, i.e. the real gap closing point, features two linear band touching points at $k_y=\pm k_{y,c}$ in the band structure as illustrated in Fig.~\ref{fig3} (a). 
Here, $k_{y,c}$ can be analytically calculated by requiring the following discriminant to be zero: $\operatorname{Disc}_{\beta_x}f(0,\beta_x,k_{y,c})=0$, where $f(E,\beta_x,k_y):=\det|E\mathbb{I}-\mathcal{H}(\beta_x,k_y)|$ is the characteristic equation.
We may further write the non-Bloch wave vector as $\beta_x=e^{\mu_x+i k_x}$, where $\mu_x$ marks its amplitude and $k_x$ marks its phase. 
Here $k_x$ must not be considered as a Bloch momentum, but rather, can be viewed as a parameter equivalent to $\phi$ in Eq.~\eqref{eq7}.
With this, we can plot the whole band structure of the system on the cylinder GBZ manifold (Fig.~\ref{fig3} (c)), which is a deformed torus parametrized by $k_x$ and $k_y$, and plot the 3D band structure as in Fig.~\ref{fig3} (d).
Having obtained $k_{y,c}$, the corresponding $k_{x,c}$ can be calculated by solving $f(0,R(\pm k_{y,c})e^{ik_{x,c}},\pm k_{y,c})=0$, where $R(k_y)$ is the radius of the cylinder GBZ given by Eq.\eqref{eq8}.
Due to the symmetry of our model (discussed in detail in Sec.~\ref{sec:classification}), we find that $k_{x,c}=\pi$ for both $\pm k_{y,c}$.
For simplicity, we define $\tilde{h}(k_x,k_y):=\tilde{H}(R(k_y)e^{ik_x},k_y)$.
By expanding $\tilde{h}(k_x,k_y)$ near $\mathbf{k}=(k_{x,c},\pm k_{y,c})$ (see Eq.~\eqref{eq:linear}), we can verify that the band closing point is a 2D Dirac point in the $(k_x,k_y)$ space.

\subsection{Topological classification of the non-Bloch Dirac points}\label{sec:classification}
Next, we explore the topological property of the non-Bloch Dirac point.
Remarkably, due to the reality of the non-Bloch spectrum, we find that the Hamiltonian can be adiabatically connected to a Hermitian Bloch band Hamiltonian with the same spectrum.
Thus, our problem reduces to the classification of Hermitian topological semimetals, which has been well-understood~\cite{Chiu2016}.

We first illustrate the mapping to a Hermitian system.
Note that the spectrum of $\tilde{h}(k_x,k_y)$ is real when $\eta{=}0$.
This indicates that we can perform a similarity transformation to map the Hamiltonian to a Hermitian matrix.
Indeed, there exists an invertible matrix $P=e^{i\frac{\pi}{4}\sigma_z}\operatorname{diag}\{1,\sqrt{\frac{t-\gamma}{t+\gamma}}\}$ such that $h(k_x,k_y):=P^{-1}\tilde{h}(k_x,k_y)P$ is Hermitian, where
\begin{align}\label{eq:Hermitianized}
    h(\mathbf{k})&{=}\mqty(0&i\sqrt{t(k_y)^2-\gamma^2}{+}ie^{ik_x}\\-i\sqrt{t(k_y)^2-\gamma^2}{-}ie^{-ik_x}&0)\nonumber\\
    &=-\sin k_x\sigma_x-\left[\cos k_x+\sqrt{t(k_y)^2-\gamma^2}\right]\sigma_y.
\end{align}
Geometrically, this similarity transformation can be viewed as flattening the cylinder GBZ manifold by homogenizing the GBZ radius $R(k_y)$.
The band structure in the $(k_x,k_y)$ space features a pair of anisotropic Dirac cones.
This becomes apparent by performing a low-energy expansion.
The gap-closing point of this Hamiltonian is at $\mathbf{k}_c^{\pm}=(k_{x,c},k^{\pm}_{y,c})=(\pi,\pm\arccos[(\sqrt{1+\gamma^2}-r)/2c])$.
Take $\mathbf{q}^{\pm}=(q_x,q^{\pm}_y):=\mathbf{k}-\mathbf{k}^{\pm}_c$ and we can expand Eq.~\eqref{eq:Hermitianized} as
\begin{equation}\label{eq:linear}
    h_{\text{eff}}(\mathbf{q}^{\pm})=q_x\sigma_x+q_y^{\pm}\sigma_y/\xi
\end{equation}
where $\xi$ is a anisotropy constant given by $\xi=\{(\gamma ^2+1)[4 c^2-(r-\sqrt{\gamma ^2+1})^2]\}^{-1/2}$.
Since $h$ is adiabatically connected to $\tilde{h}$, we only need to study the topological property of $h(\mathbf{k})$ given by Eq.~\eqref{eq:Hermitianized}, possessing time-reversal symmetry ($\mathcal{T}$), particle-hole symmetry ($\mathcal{C}$), and chiral symmetry ($\mathcal{S}$).
The symmetry operators are given by $\hat{\mathcal{T}}=\sigma_z\mathcal{K}$, $\hat{\mathcal{C}}=\mathcal{K}$, and $\hat{\mathcal{S}}=\sigma_z$ with $\hat{\mathcal{T}}^2=\hat{\mathcal{C}}^2=+1$.
Except for the above onsite symmetries, the system respects mirror symmetries $R_x$ and $R_y$.
They are given by $\hat{R_x}^{-1}h(k_x,k_y)\hat{R_x}=h(-k_x,k_y)$ with $\hat{R}_x=\sigma_y$ and $\hat{R_y}^{-1}h(k_x,k_y)\hat{R_y}=h(k_x,-k_y)$ with $\hat{R}_y=\mathbb{I}$.
To study the symmetry protection of the non-Bloch DPs, we identify all the symmetry-allowed mass terms that can be added to the Hamiltonian $h(k_x,k_y)$ without opening a gap.

\begin{table}[ht!]
\centering
    \begin{tabular}{c|ccccc}
    \toprule 
        & $\sigma_x$ & $\sigma_y$ & $\sigma_z$ & $\sin k_x\sigma_z$ & $\sin k_y\sigma_z$ \\
    \hline 
        $\hat{\mathcal{T}}=\sigma_z\mathcal{K}$ & $\times$ & \checkmark & \checkmark & $\times$ & \checkmark\\
        $\hat{\mathcal{C}}=\mathcal{K}$ & $\times$ & \checkmark & $\times$ & \checkmark & \checkmark\\
        $\hat{\mathcal{S}}=\sigma_z$ & \checkmark & \checkmark & {\color{red}$\times$} & $\times$ & {\color{red}$\times$}\\
        $R_x$ & $\times$ & \checkmark & {\color{red}$\times$} & \checkmark & {\color{red}$\times$}\\
         & gapless & gapless & {\color{red}gapped} & gapless & {\color{red}gapped}\\
    \bottomrule
    \end{tabular}
    \caption{The first row of the table lists several perturbation terms, while the first column lists the symmetries of the Hamiltonian. Each entry in the table uses checkmarks (\checkmark) and crosses ($\times$) to indicate whether the symmetry is preserved (\checkmark) or destroyed ($\times$) by a given perturbation term. The last row summarizes whether the perturbation opens a gap in the system.}
    \label{tb:1}
\end{table}

As shown in the table~\ref{tb:1}, the system remains gapless as long as either chiral symmetry ($\hat{\mathcal{S}}$) or mirror symmetry ($R_x$) is preserved. Therefore, depending on the restrictions imposed on the perturbations, the system can belong to these five symmetry classes: A$+R$, AIII, AIII$+R_+$, BDI, and BDI$+R_{+-}$.
Since the codimension is $p = d - d_{\text{FS}} = 2$ in our system, according to the classification table in Ref.~\cite{Chiu2014classification}, the non-Bloch DPs are topological and each one is characterized by a $\mathbb{Z}$ topological invariant.
These integer topological indices $\nu^{\pm}$ for the two non-Bloch DPs are given by
\begin{equation}
\nu^{\pm}=\frac{i}{2\pi}\oint_{\mathcal{C}}(Q^{\pm})^{-1}\dd Q^{\pm}\in\mathbb{Z}
\end{equation}
with $Q^{\pm}=(q_x- iq^{\pm}_y)/\sqrt{q_x^2+(q^{\pm}_y)^2}$.
Here, the integral path $\mathcal{C}$ is a loop circulating each DP.
Unless we gap the system by adding symmetry-breaking perturbation terms, this integer index remains invariant.

\begin{figure}[b]
\centering
\includegraphics[width=0.9\linewidth]{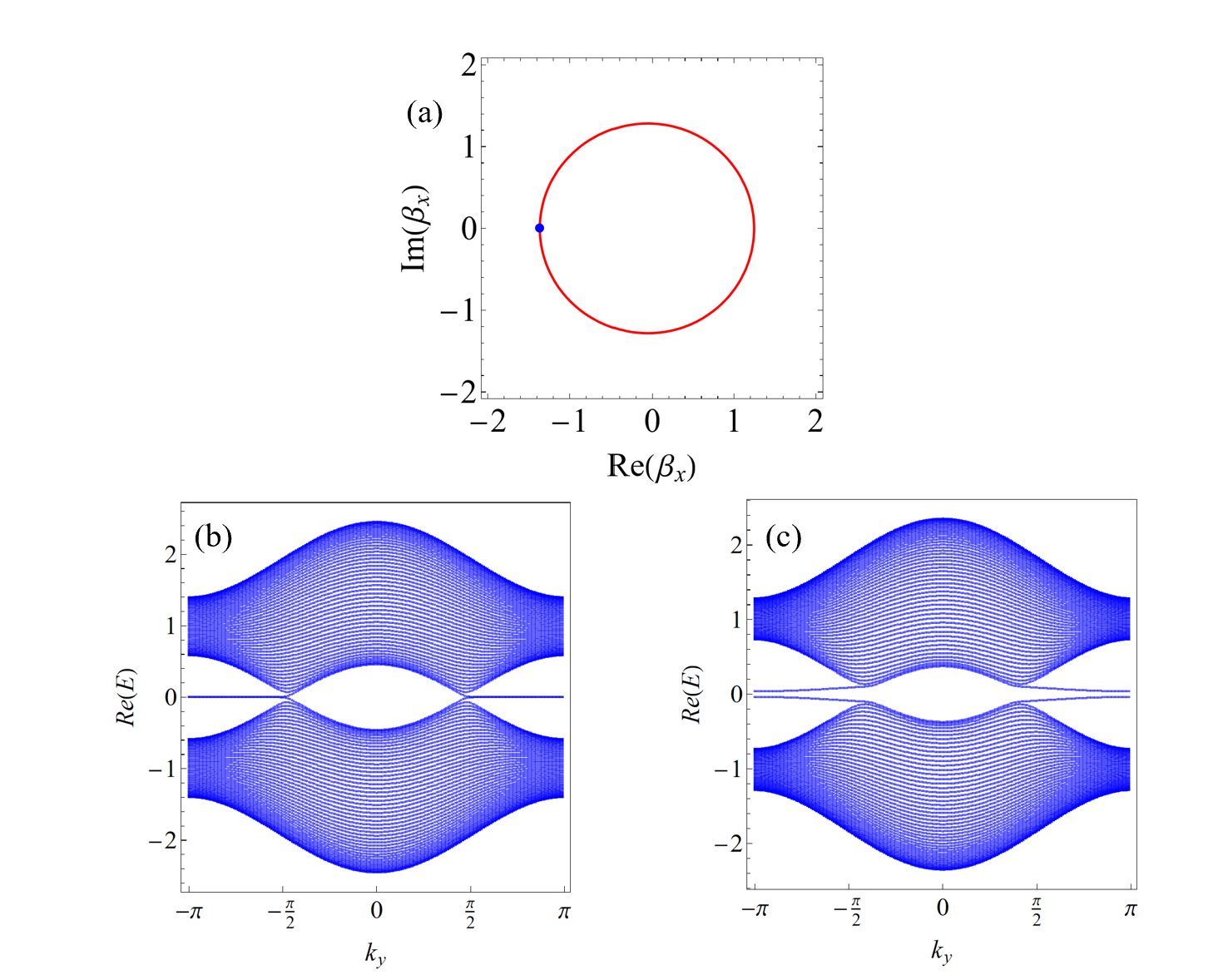}
\caption{Here we take $(r,c,\gamma,w,\eta, \delta, a)=(\frac{7}{8},\frac{1}{4},\frac{1}{4},1,0,0,\frac{1}{10})$ in (a,b) and $(r,c,\gamma,w,\eta, \delta, a)=(\frac{7}{8},\frac{1}{4},\frac{1}{4},1,0,\frac{1}{10},0)$ in (c) such that we go from phase R3 to R4 by crossing the non-Bloch DP.
In (a) the red curve is the $\operatorname{aGBZ}$ and the blue dot is the DP location. (b) shows the band structure with $\sigma_z$ perturbation and (c) with $\sigma_x$ perturbation.}
\label{fig5}
\end{figure}
\indent

\subsection{Robustness of the non-Bloch Dirac points}\label{sec:generalization}

In this section, we discuss the robustness of the non-Bloch DPs.
In the last section, we establish that the gap stays closed when symmetry-permitted perturbations are applied to the Hermitianized Hamiltonian.
Upon mapping these perturbations back to the original non-Hermitian system, we find that they retain their original form, at least locally in the vicinity of the non-Bloch DPs.
Notice that the similarity transformation $P$ is diagonal and thus commutes with $\sigma_z$, $\sigma_0$.
Therefore, the ``mass terms" for the Hamiltonian, $\sigma_z$ and $\sin k_y \sigma_z$, stay invariant after the similarity transformation.
Consequently, perturbations proportional to $\sigma_z$ or $\sin k_y\sigma_z$ adding directly to the original non-Hermitian Hamiltonian will gap the non-Bloch DPs.
For the perturbations that preserve the DPs of the Hermitian Hamiltonian $\mathcal{H}$, i.e. $\sigma_x$ and $\sigma_y$ terms, they also stay invariant locally under the $P$-transformation.
Hence, they also preserve the non-Bloch DPs when directly added to the original non-Hermitian Hamiltonian.

We further investigate the robustness of the non-Bloch DPs in the presence of long-range hopping terms. 
Let us consider a generalized version of the model in Eq.~\ref{eq2} with additional long-range hopping terms controlled by $\delta$ and $a$.
Its Hamiltonian is given by
\begin{equation}\label{eq13}
\widetilde{\mathcal{H}}_{\text{LR}}(\beta_x, k_y)=\widetilde{\mathcal{H}}+
\mqty(\dfrac{\delta}{2}(\beta _{x} + \dfrac{1}{\beta _{x}}) & a w \beta _{x}^2\\ 
\dfrac{aw}{\beta _{x}^2} & - \dfrac{\delta}{2}(\beta _{x} + \dfrac{1}{\beta _{x}})
)
\end{equation}
where $\delta$ represents the hopping strength between horizontally adjacent $A$ sublattices and $a$ represents the hopping strength between $A$ and $B$ in horizontally adjacent unit cells.
Here, we have again calculated the energy spectrum for $\widetilde{\mathcal{H}}_{\text{LR}}$ under cylindrical geometry, and we again take OBC along $x$ and take PBC along $y$. 
Using the resultant method~\cite{Yang2020}, we analytically obtain the auxiliary GBZ and corresponding energy band structure.
In Fig.~\ref{fig5} (b) and (c), we see that the off-diagonal long-range hopping controlled by $a$ preserves the non-Bloch DPs while diagonal perturbation proportional to $\delta$ opens the gap.

\begin{figure}[t]
    \centering
    \includegraphics[width=\linewidth]{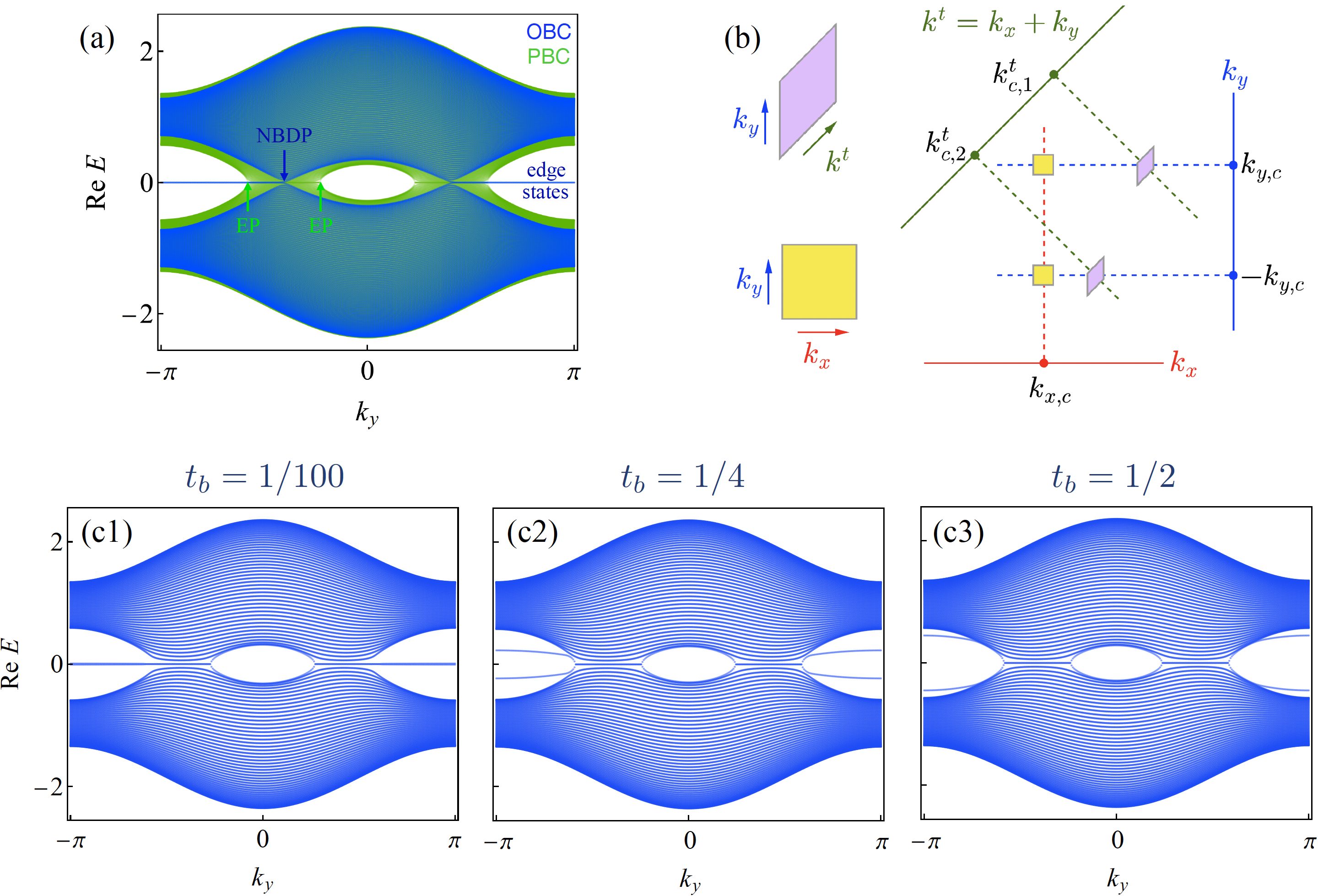}
    \caption{Crossing R3-R4 (a) real energy band structure (a), bulk-boundary correspondence for the DP system (b) and OBC to PBC transition (c, d, e). In (a) the lighter color is the OBC system and the darker color is the PBC system. The black vertical dashed lines mark the DPs in the OBC system and the pink vertical dashed lines mark the EPs in the PBC system. The DPs in the OBC system split to form a pair of EPs in the PBC system. In (b) the purple parallelogram represents the BBC under tilted geometry and the yellow square represents the BBC under square geometry. At link = 1/100 (c), the system band structure resembles the unlinked system in Fig. 3 (a). At link = 1/2, (c, f), the system matches the fully linked system in Fig.~\ref{fig3} (a).}
    \label{fig4}
\end{figure}

\section{Absence of Bulk-Boundary Correspondence in Non-Bloch Dirac Points}\label{sec:OBCPBCdistinction}

In this section, we highlight the breakdown of the conventional bulk-boundary correspondence principle in non-Hermitian semimetals. 
For Hermitian topological semimetals, the bulk Bloch band structure calculated under PBC \emph{universally} determines the topological edge states, such as surface Fermi arcs. 
The Bloch bands can be projected to any open edge of the system, where the locations of band crossing points are consistent across any boundary conditions. 
We can also predict the existence of topological edge states according to the topological number calculated from the bulk band. 
In stark contrast, we observe that the existence and locations of non-Bloch DPs are non-universal.
Notably, the non-Bloch DPs disappear and split into pairs of EPs when the boundary conditions change from OBC to PBC.
Furthermore, upon projecting the non-Bloch DPs to boundaries in different directions, we discover that their locations are incompatible with each other. 
This indicates that there is no longer universal bulk band touchings for non-Bloch semimetals. 

To demonstrate our findings, we first compare the band structure along $y$-direction under PBC and OBC, using the same parameters as in Fig.~\ref{fig3}.
As shown in Fig.~\ref{fig4} (a), each non-Bloch DP split into a pair of EPs when changing the boundary condition from OBC (blue) to PBC (green).
Furthermore, the non-Bloch DPs are very sensitive to the change of boundary condition, as we numerically calculated the transition from OBC $x$ and PBC $y$ to PBC $x$ and $y$ by adding a linking value in between $[0,1]$ of the $x$ hopping strength where $0$ is OBC $x$ and $1$ is PBC $x$. 
Even when the boundary link is very weak ($t_b=1/100$ in Fig.~\ref{fig4} (c)), the band crossing points are already widened into pairs of EPs. 
This phenomenon originates from the sensitivity of boundary topology of non-Hermitian systems~\cite{Budich2020PRL}.
We also observe the projection of Fermi Arc edge states connecting the two non-Bloch DPs (the zero-energy blue line terminating at the gap closing points as shown in Fig.~\ref{fig4} (a)), which are now governed by the non-trivial topology of the non-Bloch band structure. 

Moreover, we calculated the positions of the non-Bloch DPs for the system under square and parallelogram geometries and found that they are incompatible.
As shown in Fig.~\ref{fig4} (b), both the rectangular (yellow) and parallelogram (purple) geometries include the edge in the $y$ direction. 
Consequently, they share the same set of $k_{y,c}$ values.
For the rectangular geometry, $k_x$ is along the horizontal direction. Hence $k_{x,c}=\pi$ as discussed in Sec.~\ref{sec:DPs}.
However, for the parallelogram geometry, the tilted edges are along the $x+y$ direction.
Denoting the edge momentum by $k^t$, we obtain $k^t_{c,1}$ and $k^t_{c,2}$ which allows us to pin down the locations of the non-Bloch DPs.
The locations of the non-Bloch DPs under different boundary conditions are indicated by yellow rectangles and purple parallelograms, and they do not match.
This demonstrates that the projection doctrine in the conventional bulk-boundary correspondence is violated.
Notably, the edge-dependent nature of the non-Bloch DPs is unique to non-Hermitian systems in two and higher dimensions.

\section{Conclusion}\label{sec:conclusion}
In this work, we establish the topological classification and geometry-dependent bulk-boundary correspondence of non-Bloch Dirac points. 
To illustrate this, we use a solvable 2D stacked non-Hermitian SSH model and analytically derive the phase diagram in terms of its complex-energy point and line gaps. 
This phase diagram reveals the existence of non-Bloch Dirac points, characterized by real-spectrum gap closings that emerge under open boundary conditions but disappear with periodic boundaries. 
The reality of the spectrum near non-Bloch Dirac points allows for a local mapping to Hermitian semimetals in AZ symmetry classes, combined with an additional mirror symmetry. 
Through this mapping, we show that non-Bloch Dirac points carry an integer topological charge across different symmetry classes. 
Additionally, we find that the positions of these non-Bloch Dirac points in complex-momentum space are sensitive to boundary geometries, indicating a geometry-dependent bulk-boundary correspondence.
Our investigation elucidates the fundamental properties of 2D non-Bloch Dirac points and may illuminate pathways for further exploration and application of non-Bloch DPs. 

\acknowledgments
This research was supported by the National Science Foundation through the Materials Research Science and Engineering Center at the University of Michigan, Award No. DMR-2309029. 

\bibliographystyle{apsrev4-1}
\bibliography{refs.bib}

\end{document}